\documentclass[aps,twocolumn,prl,preprintnumbers,showpacs,%
superscriptaddress]{revtex4}
\usepackage{amsmath}
\usepackage{graphicx}
\usepackage{amsmath,amssymb,epsfig,bm}

\newcommand{\be}{\begin{equation}} \newcommand{\ee}{\end{equation}}
\newcommand{\bea}{\begin{eqnarray}} \newcommand{\eea}{\end{eqnarray}}
\newcommand{\beann}{\begin{eqnarray*}}  \newcommand{\eeann}{\end{eqnarray*}}
\newcommand{\bfig}{\begin{figure}} \newcommand{\efig}{\end{figure}}
\newcommand{\ba}{\begin{array}} \newcommand{\ea}{\end{array}}
\newcommand{\bcen}{\begin{center}} \newcommand{\ecen}{\end{center}}
\newcommand{\btab}{\begin{tabular}} \newcommand{\etab}{\end{tabular}}

\begin{document}

\title{Electric-Magnetic Duality and Topological Insulators}
\author{A.~Karch}
\affiliation{Department of Physics, University of Washington, Seattle,
WA 98195-1560, USA}

\date{July 2009}

\begin{abstract}
We work out the action of the $SL(2,\mathbb{Z})$ electric-magnetic duality group for an insulator with a non-trivial permittivity, permeability and $\theta$-angle. This theory has recently been proposed to be the correct low-energy effective action for topological insulators. As applications, we give manifestly $SL(2,\mathbb{Z})$ covariant expressions for
the Faraday rotation at orthogonal incidence at the interface of two such materials, as well as for the induced magnetic and electric charges, slightly clarifying the meaning of expressions previously derived in the literature. We also use electric-magnetic duality to find a gravitational dual for a strongly coupled version of this theory using
the AdS/CFT correspondence.
 \end{abstract}

\maketitle

\pagestyle{plain}
\emph{Introduction.}---
The Maxwell-Lagrangian of classical electromagnetism
can be modified by including a term proportional to $\theta \vec{E} \cdot \vec{B}$.
This ``axionic electrodynamics" was studied two decades ago
to describe how standard electrodynamics would be modified
in the presence of a non-trivial background for a putative
axion field. Several novel effects that would result from
such a term have been uncovered: a magnetic monopole surrounded
by a small bubble of vacuum with no axion but an otherwise constant
axion field throughout space would pick up a non-trivial electric charge
\cite{Wilczek:1987mv}
giving a nice realization of the Witten effect \cite{Witten:1979ey}. Magnetic
charges induce electric mirror charges and vice versa in the
presence of a planar domain wall across which $\theta$ jumps \cite{Sikivie:1984yz} . Reflection off such a domain wall
induces a non-trivial rotation of the polarization of the field
\cite{Huang:1985tt}.

Recently interest in this theory has seen a sudden revival as it was
argued in \cite{Qi:2008ew} that it is also the low energy effective action
for so called ``topological insulators". In this context one fascinating
property of an interface with a non-trivial jump in $\theta$ has
been emphasized \cite{Xiao-LiangQi02272009}: an electric charge
close to the interface induces a mirror {\it magnetic} charge. While this was originally derived in
\cite{Sikivie:1984yz}, \cite{Xiao-LiangQi02272009} generalized the calculation to materials with non-trivial
permittivity and permeability and of course emphasized the potential experimental relevance in the context
of topological insulators.

Motivated by this observation we take a look at the
action of electric-magnetic duality in this theory.
$U(1)$ gauge theories with non-zero $\theta$ are well known
to exhibit an $SL(2,\mathbb{Z})$ duality group which strongly
constrains the quantum physics, see e.g. \cite{Witten:1995gf}.
In this letter we work out the action of $SL(2,\mathbb{Z})$
on an insulator with non-trivial permittivity, permeability
and $\theta$ angle. This allows us to write the formula
for the induced mirror charge as well as the one for
the Faraday effect in a manifestly $SL(2,\mathbb{Z})$ covariant
form. Along the way we also clarify the role of the mirror charges
calculated in \cite{Xiao-LiangQi02272009}. While the mirror charges
calculated in \cite{Xiao-LiangQi02272009} allow for a correct determination
of the electric and magnetic fields that are sourced by the charge
next to the interface, what appears to
be an electric mirror charge in that work
is in fact a dyonic mirror charge.

Electric-magnetic duality maps
interfaces between two ordinary insulators
into an interface between an ordinary and a topological insulator
in the dual theory. We use this fact to construct gravitational
duals for topological insulators in strongly coupled ${\cal N}=4$ SYM
via the AdS/CFT correspondence.

\vspace{.3cm}

\emph{Electric-Magnetic Duality in matter.}---
Let us begin with standard Maxwell's equations for
electromagnetism in matter in Gaussian units.
They derive from a Lagrangian
\be
S_0 = \int d^3x dt L_0=
\frac{1}{8 \pi} \int d^3x dt \, \left (\epsilon \vec{E}^2
 - \frac{1}{\mu} \vec{B}^2 \right ).
\ee
$\epsilon$ and $\mu$ are the permittivity and permeability of the medium.
The corresponding vacuum quantities $\epsilon_0$, $\mu_0$
in Gaussian units are set to 1, see e.g. \cite{jackson}. We also set $c_0$, the speed of light in vacuum, to 1.
In matter electromagnetic waves propagate with a velocity of light
that is $c = (\mu \epsilon)^{-1/2}$.
With these conventions Maxwell's equations read
\be
\begin{array}{lcllrlcl}
\vec{\nabla} \cdot \vec{D} &=& 4 \pi \rho_e, &\hspace{.4cm}&& \hspace{-.1cm}
\vec{\nabla} \times \vec{H} &=&
\frac{\partial \vec{D}}{\partial t} + 4 \pi \vec{j}_e \\
\vec{\nabla} \cdot \vec{B} &=&  4 \pi \rho_m, &\hspace{.4cm}&-& \hspace{-.1cm}
\vec{\nabla} \times  \vec{E} &=&
\frac{\partial \vec{B}}{\partial t} + 4 \pi \vec{j}_m
\end{array}
\ee
with the constitutive relations
\be \vec{D}=
4 \pi \frac{\delta L_0}{\delta \vec{E}} =\epsilon \vec{E}, \hspace{.4cm} \vec{H} =
- 4 \pi \frac{\delta L_0} {\delta \vec{B}}= \frac{\vec{B}}{\mu}.
\ee
$\rho_{e,m}$ and $\vec{j}_{e,m}$ are the electric (magnetic) charge and
current densities respectively.
In vacuum these equations are invariant under duality rotations
\be
\begin{pmatrix}
 \vec{D} \\ N\vec{ B}
\end{pmatrix}
= {\cal R}_{\xi}
\begin{pmatrix}
 \vec{D}' \\ N\vec{ B}'
\end{pmatrix},
\hspace{.5cm}
\begin{pmatrix}
N \vec{E} \\ \vec{H}
\end{pmatrix}
= {\cal R}_{\xi}
\begin{pmatrix}
N \vec{E}' \\ \vec{H}'
\end{pmatrix}
\ee
where ${\cal R}_{\xi}= \begin{pmatrix} \cos \xi & \sin \xi \\
- \sin \xi & \cos \xi \end{pmatrix}$ is a standard rotation matrix
and $N$ an arbitrary normalization constant.
In the presence of charges this duality is still a symmetry of the classical
equations of motion as long as the charges transform under the duality as well
\be
\begin{pmatrix}
\rho_e \\ N \rho_m
\end{pmatrix}
= {\cal R}_{\xi}
\begin{pmatrix}
\rho_e' \\ N \rho_m'
\end{pmatrix},
\hspace{.5cm}
\begin{pmatrix}
 \vec{j}_e \\ N\vec{j}_m
\end{pmatrix}
= {\cal R}_{\xi}
\begin{pmatrix}
 \vec{j}_e' \\ N\vec{j}_m'
\end{pmatrix}
\ee
Quantum mechanically the total electric charge $q_e=\frac{1}{4 \pi} \int d\vec{S} \cdot \vec{D}$
($S$ being any closed surface) has to be an integer multiple of the electron charge $e$, whereas the magnetic charge
$q_m=\frac{1}{4 \pi} \int d\vec{S} \cdot \vec{B}$
by the Dirac quantization condition should be an integer multiple of $g=\frac{e}{2 \alpha}$
where $\alpha = \frac{e^2}{\hbar c}$ is the fine-structure constant. Only the
special case of a duality transformation with $\xi=\pi/2$ and $N= 2 \alpha$ is consistent with leaving
this requirement invariant.
This transformation typically referred to as the S-generator of electric-magnetic duality.
In addition to its action on the fields it only leaves the constitutive relations
invariant if we exchange $\epsilon/N$ and $N \mu$.
The speed of light $(\mu \epsilon)^{-1/2}$
is invariant under S-duality.

For a topological insulator (or electromagnetism in the presence
of a constant axion field) the effective action contains a 2nd term. That is
$S=S_0 - S_{\theta}$ where
\be
S_{\theta} =  \frac{\theta}{2 \pi} \frac{\alpha}{16 \pi}
\int d^3x dt \, \epsilon^{\mu \nu \rho \sigma} F_{\mu \nu} F_{\rho \sigma}
= \frac{\theta}{2 \pi} \frac{\alpha}{2 \pi}
\int d^3x dt \, \vec{E} \cdot \vec{B}.
\ee
Maxwell's equations are untouched by this addition. However the
constitutive relations are modified to
\begin{eqnarray}
\nonumber
\vec{D} &=& \epsilon \vec{E} -  \frac{\theta}{2 \pi} \, (2 \alpha \vec{B}) \\
\vec{H} &=& \frac{\vec{B}}{\mu} + \frac{\theta}{2 \pi} (2 \alpha \vec{E})
\end{eqnarray}
Classically these equations are invariant under shifts of $\theta$ by any constant,
$\theta = \theta' +C$ together with
\be
\nonumber
\begin{pmatrix}
 \vec{D} \\  2 \alpha \vec{ B}
\end{pmatrix}
= \Lambda
\begin{pmatrix}
 \vec{D}' \\ 2 \alpha \vec{ B}'
\end{pmatrix},
\hspace{.3cm}
\begin{pmatrix}
2 \alpha \vec{E} \\ \vec{H}
\end{pmatrix}
= (\Lambda^T)^{-1}
\begin{pmatrix}
2 \alpha \vec{E}' \\ \vec{H}'
\end{pmatrix}
\ee
\be
\label{duality}
\begin{pmatrix}
\rho_e \\ 2 \alpha \rho_m
\end{pmatrix}
= \Lambda
\begin{pmatrix}
\rho_e' \\ 2 \alpha \rho_m'
\end{pmatrix},
\hspace{.5cm}
\begin{pmatrix}
 \vec{j}_e \\ 2 \alpha \vec{j}_m
\end{pmatrix}
= \Lambda
\begin{pmatrix}
 \vec{j}_e' \\ 2 \alpha \vec{j}_m'
\end{pmatrix}
\ee
where $\Lambda = \begin{pmatrix} 1 & -\frac{C}{2 \pi} \\ 0 & 1 \end{pmatrix}$ and
hence $(\Lambda^T)^{-1} =\begin{pmatrix} 1 & 0 \\ \frac{C}{2 \pi}  & 1
\end{pmatrix}$.
Classically we can always use this symmetry to set $\theta$ to
zero.
Quantum mechanically however shifts in $\theta$ are only a symmetry if
$C$ is an integer multiple of
$2 \pi$. While shifts in $\theta$ do leave the equations of motion invariant
they change the action and hence the weight $e^{ i S/\hbar}$
in the path integral.
As long as electric and magnetic
fluxes are properly quantized $S_{\theta}/\hbar$ however is an integer
multiple of $\theta$ (see e.g. \cite{Witten:1995gf}), and so shifts of $\theta$ by
integer multiples of $2 \pi$ leave the path integral invariant.
Consequently values of $\theta$ between 0 and $2 \pi$ are physically distinct. Only $\theta=0$
and $\theta=\pi$ give a time-reversal symmetric theory. Time reversal
takes $\theta$ into $-\theta$, so $\theta=0$ is
time reversal invariant as it stands, whereas $\theta=\pi$ is invariant after
a shift of $\theta$ by $2 \pi.$ A shift of $\theta$ by $2 \pi$ is typically
referred to as the T-generator of electric-magnetic duality.

The full quantum mechanical duality group is obtained by repeated
application of the T- and S-generators. The two together
generate an $SL(2,\mathbb{Z})$ symmetry which
acts on the fields as in eq. (\ref{duality}) with a general
matrix $\Lambda = \begin{pmatrix} {\tt a} & {\tt b} \\ {\tt c} & {\tt d} \end{pmatrix}$
with integers ${\tt a}$, ${\tt b}$, ${\tt c}$ and ${\tt d}$ satisfying
${\tt a}{\tt d}-{\tt b}{\tt c}=1$.

To determine
the action of a general $SL(2,\mathbb{Z})$
transformation on the three parameters in the constitutive relation
$\epsilon$, $\mu$ and $\theta$ note that the constitutive relation can
be written in the compact form
\be
\label{constitutive}
\begin{pmatrix} \vec{D} \\ 2 \alpha \vec{B} \end{pmatrix} = {\cal M}
\begin{pmatrix}  2 \alpha \vec{E} \\ \vec{H} \end{pmatrix}
\ee
with
\be
{\cal M} = \frac{1}{c} \frac{2 \alpha}{c \epsilon}
\begin{pmatrix}  \frac{\theta^2}{4 \pi^2}
+ \left ( \frac{c \epsilon}{2 \alpha} \right )^2 & -  \frac{\theta}{2 \pi} \\
- \frac{\theta}{2 \pi} & 1 \end{pmatrix}
\ee
where we have replaced $1/\mu = c^2 \, \epsilon$.
The transformation eq. (\ref{duality})
is a symmetry as long as ${\cal M}$ transforms as
\be
\label{mtrafo}
{\cal M} = \Lambda {\cal M}' \Lambda^T.
\ee
$1/c^2 =\det({\cal M})$ is duality invariant.
This transformation on the matrix ${\cal M}$ can
also be written as a transformation
$\tau ' = \frac{a \tau + b}{c \tau +d}$ for the complexified
parameter $\tau = \frac{\theta}{2 \pi} + i \frac{c \epsilon}{2 \alpha}$.
The duality transformation of $\mu$ then follows from the invariance
of the speed of light, $1/\mu' = c^2\, \epsilon'$.

\vspace{.3cm}

\emph{Duality Covariant Expression for Mirror Charges.}---
With the formalism we have set up in the previous section
it is easy to redo the calculation of \cite{Xiao-LiangQi02272009}
in a manifestly duality covariant fashion. The goal is to calculate
the electric and magnetic fields resulting from a single test charge
with charge $\vec{q}=\begin{pmatrix} q_e \\ 2 \alpha q_m
\end{pmatrix}$ at a distance $d$ from a planar
interface (which we take to be the $z=0$ plane)
between two materials characterized by ${\cal M}_{1,2}$, that
is with different $\mu_i$, $\epsilon_i$ and $\theta_i$ ($i=1,2$). Of
particular interest is the case with $\theta_1=0$ and $\theta_2 = \pi$,
the interface between an ordinary and a topological conductor. But here our goal
is to work out the generic case.

For a static electromagnetism problem in the absence of currents the easiest
way to proceed is to introduce electric and magnetic potentials
$\Phi_{e,m}$ with
$\vec{D} = - \vec{\nabla} \Phi_e$ and $\vec{B} = - \vec{\nabla} \Phi_m$.
Above the interface they are given by
\be \Phi_e^I = \frac{q_e}{R_1} + \frac{q^{(2)}_e}{R_2},
\hspace{.3cm}
\Phi_m^I = \frac{q_m}{R_1} + \frac{q^{(2)}_m}{R_2}
\ee
and below by
\be \Phi_e^{II} = \frac{q_e}{R_1} + \frac{q^{(1)}_e}{R_1},
\hspace{.3cm}
\Phi_m^{II} = \frac{q_m}{R_1} + \frac{q^{(1)}_m}{R_1}
\ee
where $q_{e,m}^{(1,2)}$ are the mirror charges locate a distance $d$ above
(for $q_{e,m}^{(1)}$) or below (for $q_{e,m}^{(2)}$) the interface. $R_1^2=x^2+y^2+(d-z)^2$
and $R_2^2 = x^2 + y^2 + (d+z)^2$. Maxwell's equations
in the absence of surface currents or charges as usual demand continuity of
the perpendicular components of
$\begin{pmatrix} \vec{D} \\ 2 \alpha \vec{B} \end{pmatrix}$ and
the parallel components of $\begin{pmatrix} 2 \alpha \vec{E} \\  \vec{H} \end{pmatrix}$
or in other words
\be
\vec{q}^{(1)} = - \vec{q}^{(2)}, \hspace{.3cm}
\left ( \vec{q} +\vec{q}^{(2)} \right ) =
{\cal T} \left ( \vec{q} +\vec{q}^{(1)} \right ).
\ee
where ${\cal T} = {\cal M}_1 {\cal M}_2^{-1}$. As
the transformation properties of ${\cal M}$ imply that
\be
{\cal T}  = \Lambda {\cal T}' \Lambda^{-1}
\ee
these expressions are manifestly $SL(2,\mathbb{Z})$ covariant.
They can be solved by matrix multiplication
\be
\vec{q}^{(2)} = - \vec{q}^{(1)} = ({\cal T}+1)^{-1} ({\cal T}-1) \, \vec{q}. \ee
For the magnetic charges $q_m^{(1,2)}$ this expression is in perfect
agreement with the corresponding expressions for $g_{1,2}$ in
\cite{Xiao-LiangQi02272009}. For the electric charges there seems to
be a disagreement with the values $q_{1,2}$ obtained in
\cite{Xiao-LiangQi02272009}. In particular, our mirror electric charges
are equal opposite whereas theirs are equal. This apparent discrepancy can quickly
be resolved by noting that $q_{1,2}$ in \cite{Xiao-LiangQi02272009}
are not actually the induced electric mirror charges.
Proper electric charges (where physical charges are quantized in units of the electron
charge $e$) are the sources of $\vec{D}$ flux.
The $q_i$ of
that work were introduced as giving the sources of $\vec{E}$ flux
instead.
Due to the modified
constitutive relations in the presence of the $\theta$ terms
these two differ by a multiple of the magnetic charge.
As the mirror charges are simply
a useful technical construct to obtain the correct electric
and magnetic fields (and aren't quantized in any case) this does not
in any way alter the exciting findings of \cite{Xiao-LiangQi02272009}
which were mostly concerned with the induced magnetic charge. The proper
mirror electric charges are related
to
the quantities calculated
in \cite{Xiao-LiangQi02272009} by $(q+q_e^{(1)} +
q_m^{(1)}
\alpha \theta_2/\pi
)/\epsilon_2 = (q/\epsilon_1+ q_1)$
and $(q_e^{(2)} +
q_m^{(2)}
\alpha \theta_2/\pi
)/\epsilon_1 = q_2$.
After this substitution their results are seen to be in full agreement with
the expressions derived here.

\vspace{.3cm}

\emph{Faraday effect in reflection.}--
Our tools can also be used to calculate
reflection and refraction of light on interfaces between two materials
with different $\epsilon$, $\mu$ and $\theta$ in a
duality covariant framework.
Snell's law is unmodified in the presence of
a jump in $\theta$, so the only task is to calculate the reflected and
transmitted amplitudes.
The full expressions
have been worked out before in
\cite{Obukhov:2005kh} and all we want to do here is to exhibit
$SL(2,\mathbb{Z})$ covariance. This is most easily done in the case of
orthogonal incidence to which we'll restrict ourselves from now on.
In the presence of a non-trivial
$\theta$ angle Maxwell's equations still allow propagating wave solutions
with $\omega = c k$, all fields orthogonal to the direction of propagation,
$\hat{k}$, and
\be
\label{amprelation}
\begin{pmatrix} 2 \alpha \vec{E} \\  \vec{H} \end{pmatrix} = c \hat{k} \times
\begin{pmatrix} 0&-1 \\1&0 \end{pmatrix}
\begin{pmatrix}   \vec{D} \\ 2 \alpha \vec{B} \end{pmatrix}.
\ee
That is, $\vec{E}$ is still orthogonal to $\vec{B}$ and so is $\vec{H}$ to $\vec{D}$.
However the 2-bein defined by $\hat{H}$,$\hat{D}$ is no longer aligned with
the $\hat{E}$, $\hat{B}$ 2-bein. For orthogonal incidence all fields are
parallel to the interface and the only boundary condition is
continuity of $\vec{\cal E} = \begin{pmatrix} 2 \alpha \vec{E}  \\ \vec{H} \end{pmatrix}$.
So we directly get
$ \vec{\cal E}_{in} + \vec{\cal E}_R = \vec{\cal E}_T$
where the subscripts $in$, $R$ and $T$ stand for the incoming,
reflected and transmitted fields respectively. As $\vec{E}$ and $\vec{H}$
are not independent for a wave solution, we get a second equation from
eq. \ref{amprelation}. Together with our constitutive relation
eq. \ref{constitutive} and the fact that
for orthogonal incidence we simply have $\hat{k}_{in} = \hat{k}_T
= - \hat{k}_R$. we get for a wave incoming from medium 2
\be
c_2 {\cal M}_2 (\vec{\cal E}_{in} - \vec{\cal E}_{R}) = c_1 {\cal M}_1
\vec{\cal E}_{T} = c_1 {\cal M}_1 (\vec{\cal E}_{in} + \vec{\cal E}_{R}).
\ee
As for the mirror charge calculation above, this now can be solved by
matrix multiplication in terms of $\tilde{\cal T}
= \frac{c_1}{c_2} {\cal M}_2^{-1} {\cal M}_1$
\be
\vec{\cal E}_R = (1 + \tilde{\cal T})^{-1} (1 - \tilde{\cal T}) \vec{\cal E}_{in}
\ee
which is in nice agreement with the formulas
in \cite{Obukhov:2005kh}. Note in particular that for an incoming wave
with $$\vec{\cal E}_{in} = \begin{pmatrix} 2 \alpha \vec{E} \\ c_2 \epsilon_2 \hat{k}
\times \vec{E} + 2 \alpha \theta_2/(2 \pi) \vec{E} \end{pmatrix}$$
the reflected wave has a component in the $\hat{k} \times \vec{E}$ direction.
That is the polarization got rotated by an angle $\beta$ with
\be
\tan(\beta)
= 4 \alpha \frac{(\theta_2 - \theta_1)}{2 \pi} \frac{ \mu_1 \sqrt{\epsilon_2 \mu_2}}{
\epsilon_2 \mu_1 - \mu_2 \epsilon_1} + {\cal O}(\alpha^2).
\ee
Like the induced magnetic mirror charge, this non-trivial Faraday
angle is a direct consequence of having a jump in $\theta$.
This formula is in perfect agreement with  \cite{Qi:2008ew,
Sikivie:1984yz}.

\vspace{.3cm}

\emph{Dynamical Axion.}-- Another interesting
property of topological insulators recently uncovered
in \cite{li-2009} is that once the $\theta$ angle is promoted
to a dynamical axion field, for example by considering
anti-ferromagnetic long range order, a new mode called an ``axionic
polariton" appears in the low energy spectrum. In the presence of a background
magnetic field the axionic polariton is a coupled mode of light and axion
field. Its dispersion relation exhibits two branches, separated by
a gap whose magnitude can be dialed by dialing the background magnetic
field. It is interesting to ask if this phenomenon can also be
described in our $SL(2,\mathbb{Z})$ covariant framework. As $SL(2,\mathbb{Z})$
maps $\theta$ into $\epsilon$ promoting just the axion to a dynamical field
is inconsistent with duality. Duality can be preserved by
also promoting $\epsilon$ and with it $\mu^{-1}=c^2 \epsilon$
into a dynamical field, the dilaton.
One can ask whether
the diable gap persists
in the $SL(2,\mathbb{Z})$ invariant system with a dynamical axion/dilaton pair.
Unfortunately with a dynamic dilaton a constant background magnetic field
is no longer a stationary background solution. The $\vec{B}^2/\mu$ term
in the Lagrangian acts as a source for the dilaton. A constant
$B$ field at time $t=0$ will lead to a time-dependent dilaton solution that
runs away to weak coupling.

\vspace{.3cm}

\emph{AdS/CFT realization via Janus.}---
Using $SL(2,\mathbb{Z})$ we can also construct
a strongly coupled analog of the interface between topological
and regular insulator using the AdS/CFT correspondence. Note that
an $SL(2,\mathbb{Z})$ duality transformation takes a theory in which only $\epsilon$ and $\mu$ jump across
the interface into one that also has a non-trivial jump in $\theta$. There
is a well known example of a AdS/CFT setup that corresponds to an
interface with jump in $\mu$ and $\epsilon$,
the Janus solution \cite{Bak:2003jk}. The analog of electromagnetism here
is ${\cal N}=4$ SYM, a theory that also mediates a Coulomb force between charge carriers.
The theory has a dual supergravity description in the limit that
the analog of $\alpha$ is very large. In addition,
the number of degrees of freedom $N_c$ is large in the supergravity
limit so that the theory is effectively classically. The Janus solution describes
an interface in this theory across which $\epsilon$ and $\mu$ jump,
but with a uniform speed of light, that is $c_1 = c_2$ \cite{Bak:2003jk,Clark:2004sb}.
The dual supergravity description has the full $SL(2,\mathbb{R})$ duality of the classical
theory. New axionic Janus solutions with a non-trivial jump in $\theta$
can be produced by applying an $SL(2,\mathbb{R})$ transformation on the original Janus solution.
One obtains a supergravity solution in which, in addition to the dilaton that
already had a non-trivial profile in the original Janus solution, also the axion
pseudoscalar is turned on. It was shown in \cite{D'Hoker:2006uu} that
the most general supergravity solution
with only dilaton, axion and metric turned on and preserving the same
global symmetry as
the original Janus solution can be obtained as an $SL(2,\mathbb{R})$ transform
of Janus.

This way we can find supergravity solutions realizing any change in $\theta$ we wish.
As the speed of light is duality invariant, all of these axionic
Janus solutions have the property that $c_1=c_2$. It is easy to see why
the Janus Ansatz does not allow for interfaces across which $c$ jumps: in the
Janus construction one is looking for solutions that preserve the
2+1 dimensional relativistic conformal symmetry, $SO(3,2)$. This includes
the 2+1d Lorentz group as a subgroup. In a theory with a varying speed of light
this is not a good symmetry. To look for generalized Janus solutions realizing
an interface with a generic jump in $\mu$, $\epsilon$ and $\theta$ one has to
look for metrics in which $g_{tt}/g_{xx}$ has a non-trivial dependence on the radial
coordinate. Unlike in the Janus case, where the $SO(3,2)$ symmetry assured
that Einstein's equations boil down to simple ODEs, these
generalized Janus solutions have to be solutions to PDEs in two variables.
It would be nice to see if they can be constructed explicitly as they would be new examples of singularity free solutions of type IIB supergravity. For the purpose of realizing interfaces with a non-trivial jump in
$\theta$ it is sufficient to work with $c_1=c_2$.

\vspace{.3cm}

\emph{Conclusions.}---
In this letter we've laid out the action of the $SL(2,\mathbb{Z})$
electric-magnetic duality
group in the effective theory proposed to describe topological insulators.
We demonstrated that several classical phenomena
in these unusual
materials,
like
an induced magnetic mirror charge and a non-trivial Faraday
effect,
can be compactly worked out in a manifestly $SL(2,\mathbb{Z})$ covariant
way. We also used the fact that this duality relates the interface
between two ordinary insulators to an interface between a topological and
an ordinary insulator to identify the supergravity description of a
strongly coupled analog of topological insulators.
We hope that in the future electric-magnetic duality will
be useful to unravel the properties of these interesting materials.

\vspace{.3cm}

\emph{Acknowledgments.}---%
Thanks to S.~Hartnoll, C.~Herzog, K.~Jensen, D.~Son,
and especially L.~Yaffe for useful discussions. This work was supported in part by  U.S.\ Department
    of Energy under Grant No.~DE-FG02-96ER40956. I'd also like to thank the KITP in Santa Barbara for hospitality during the final stages of this work. There this research was supported in part by the National Science Foundation under Grant No. PHY05-51164.

\bibliographystyle{apsrev}
\bibliography{topinsulator}

\begin{thebibliography}{12}
\expandafter\ifx\csname natexlab\endcsname\relax\def\natexlab#1{#1}\fi
\expandafter\ifx\csname bibnamefont\endcsname\relax
  \def\bibnamefont#1{#1}\fi
\expandafter\ifx\csname bibfnamefont\endcsname\relax
  \def\bibfnamefont#1{#1}\fi
\expandafter\ifx\csname citenamefont\endcsname\relax
  \def\citenamefont#1{#1}\fi
\expandafter\ifx\csname url\endcsname\relax
  \def\url#1{\texttt{#1}}\fi
\expandafter\ifx\csname urlprefix\endcsname\relax\def\urlprefix{URL }\fi
\providecommand{\bibinfo}[2]{#2}
\providecommand{\eprint}[2][]{\url{#2}}

\bibitem[{\citenamefont{Wilczek}(1987)}]{Wilczek:1987mv}
\bibinfo{author}{\bibfnamefont{F.}~\bibnamefont{Wilczek}},
  \bibinfo{journal}{Phys. Rev. Lett.} \textbf{\bibinfo{volume}{58}},
  \bibinfo{pages}{1799} (\bibinfo{year}{1987}).

\bibitem[{\citenamefont{Witten}(1979)}]{Witten:1979ey}
\bibinfo{author}{\bibfnamefont{E.}~\bibnamefont{Witten}},
  \bibinfo{journal}{Phys. Lett.} \textbf{\bibinfo{volume}{B86}},
  \bibinfo{pages}{283} (\bibinfo{year}{1979}).

\bibitem[{\citenamefont{Sikivie}(1984)}]{Sikivie:1984yz}
\bibinfo{author}{\bibfnamefont{P.}~\bibnamefont{Sikivie}},
  \bibinfo{journal}{Phys. Lett.} \textbf{\bibinfo{volume}{B137}},
  \bibinfo{pages}{353} (\bibinfo{year}{1984}).

\bibitem[{\citenamefont{Huang and Sikivie}(1985)}]{Huang:1985tt}
\bibinfo{author}{\bibfnamefont{M.~C.} \bibnamefont{Huang}} \bibnamefont{and}
  \bibinfo{author}{\bibfnamefont{P.}~\bibnamefont{Sikivie}},
  \bibinfo{journal}{Phys. Rev.} \textbf{\bibinfo{volume}{D32}},
  \bibinfo{pages}{1560} (\bibinfo{year}{1985}).

\bibitem[{\citenamefont{Qi et~al.}(2008)\citenamefont{Qi, Hughes, and
  Zhang}}]{Qi:2008ew}
\bibinfo{author}{\bibfnamefont{X.-L.} \bibnamefont{Qi}},
  \bibinfo{author}{\bibfnamefont{T.}~\bibnamefont{Hughes}}, \bibnamefont{and}
  \bibinfo{author}{\bibfnamefont{S.-C.} \bibnamefont{Zhang}},
  \bibinfo{journal}{Phys. Rev.} \textbf{\bibinfo{volume}{B78}},
  \bibinfo{pages}{195424} (\bibinfo{year}{2008}), \eprint{0802.3537}.

\bibitem[{\citenamefont{Qi et~al.}(2009)\citenamefont{Qi, Li, Zang, and
  Zhang}}]{Xiao-LiangQi02272009}
\bibinfo{author}{\bibfnamefont{X.-L.} \bibnamefont{Qi}},
  \bibinfo{author}{\bibfnamefont{R.}~\bibnamefont{Li}},
  \bibinfo{author}{\bibfnamefont{J.}~\bibnamefont{Zang}}, \bibnamefont{and}
  \bibinfo{author}{\bibfnamefont{S.-C.} \bibnamefont{Zhang}},
  \bibinfo{journal}{Science} \textbf{\bibinfo{volume}{323}},
  \bibinfo{pages}{1184} (\bibinfo{year}{2009}),
  \eprint{http://www.sciencemag.org/cgi/reprint/323/5918/1184.pdf}.

\bibitem[{\citenamefont{Witten}(1995)}]{Witten:1995gf}
\bibinfo{author}{\bibfnamefont{E.}~\bibnamefont{Witten}},
  \bibinfo{journal}{Selecta Math.} \textbf{\bibinfo{volume}{1}},
  \bibinfo{pages}{383} (\bibinfo{year}{1995}), \eprint{hep-th/9505186}.

\bibitem[{\citenamefont{Jackson}(1998)}]{jackson}
\bibinfo{author}{\bibfnamefont{J.~D.} \bibnamefont{Jackson}},
  \emph{\bibinfo{title}{Classical Electrodynamics}}
  (\bibinfo{publisher}{Wiley}, \bibinfo{year}{1998}), \bibinfo{edition}{3rd}
  ed., ISBN \bibinfo{isbn}{047130932X}.

\bibitem[{\citenamefont{Obukhov and Hehl}(2005)}]{Obukhov:2005kh}
\bibinfo{author}{\bibfnamefont{Y.~N.} \bibnamefont{Obukhov}} \bibnamefont{and}
  \bibinfo{author}{\bibfnamefont{F.~W.} \bibnamefont{Hehl}},
  \bibinfo{journal}{Phys. Lett.} \textbf{\bibinfo{volume}{A341}},
  \bibinfo{pages}{357} (\bibinfo{year}{2005}), \eprint{physics/0504172}.

\bibitem[{\citenamefont{Bak et~al.}(2003)\citenamefont{Bak, Gutperle, and
  Hirano}}]{Bak:2003jk}
\bibinfo{author}{\bibfnamefont{D.}~\bibnamefont{Bak}},
  \bibinfo{author}{\bibfnamefont{M.}~\bibnamefont{Gutperle}}, \bibnamefont{and}
  \bibinfo{author}{\bibfnamefont{S.}~\bibnamefont{Hirano}},
  \bibinfo{journal}{JHEP} \textbf{\bibinfo{volume}{05}}, \bibinfo{pages}{072}
  (\bibinfo{year}{2003}), \eprint{hep-th/0304129}.

\bibitem[{\citenamefont{Clark et~al.}(2005)\citenamefont{Clark, Freedman,
  Karch, and Schnabl}}]{Clark:2004sb}
\bibinfo{author}{\bibfnamefont{A.~B.} \bibnamefont{Clark}},
  \bibinfo{author}{\bibfnamefont{D.~Z.} \bibnamefont{Freedman}},
  \bibinfo{author}{\bibfnamefont{A.}~\bibnamefont{Karch}}, \bibnamefont{and}
  \bibinfo{author}{\bibfnamefont{M.}~\bibnamefont{Schnabl}},
  \bibinfo{journal}{Phys. Rev.} \textbf{\bibinfo{volume}{D71}},
  \bibinfo{pages}{066003} (\bibinfo{year}{2005}), \eprint{hep-th/0407073}.

\bibitem[{\citenamefont{D'Hoker et~al.}(2006)\citenamefont{D'Hoker, Estes, and
  Gutperle}}]{D'Hoker:2006uu}
\bibinfo{author}{\bibfnamefont{E.}~\bibnamefont{D'Hoker}},
  \bibinfo{author}{\bibfnamefont{J.}~\bibnamefont{Estes}}, \bibnamefont{and}
  \bibinfo{author}{\bibfnamefont{M.}~\bibnamefont{Gutperle}},
  \bibinfo{journal}{Nucl. Phys.} \textbf{\bibinfo{volume}{B757}},
  \bibinfo{pages}{79} (\bibinfo{year}{2006}), \eprint{hep-th/0603012}.

\end{thebibliography}

\end{document}